\documentclass[namedreferences,hyperref,optionalrh,solaromanenum]{spr-sola}
\usepackage{natbib}
\usepackage[normalem]{ulem}
\usepackage{graphicx}        
\usepackage{color}           

\usepackage{amsmath,amssymb}
\usepackage{bm}
\usepackage{breakurl}        

\chardef\us=`\_

\begin{document}

\begin{article}
\begin{opening}

\title{Combined Surface Flux Transport and Helioseismic Far-side Active Region Model (FARM)}

\author[addressref=MPS,corref,email={yangd@mps.mpg.de}]{\inits{D.Y.}\fnm{Dan}~\lnm{Yang}\orcid{0000-0001-7570-1299}}

\author[addressref={UH,MPS}]{\inits{S.G.H.}\fnm{Stephan G.}~\lnm{Heinemann}\orcid{0000-0002-2655-2108}}%

\author[addressref={MPS}]{\inits{R.H.C.}\fnm{Robert H.}~\lnm{Cameron}\orcid{0000-0001-9474-8447}}

\author[addressref={MPS,UG,UAD}]{\inits{L.G.}\fnm{Laurent}~\lnm{Gizon}\orcid{0000-0001-7696-8665}}

\address[id=MPS]{Max-Planck-Institut für Sonnensystemforschung, Justus-von-Liebig-Weg 3, 37077 G\"ottingen, Germany}

\address[id=UH]{Department of Physics, University of Helsinki, P.O. Box 64, 00014, Helsinki, Finland}

\address[id=UG]{Institut f\"ur Astrophysik, Georg-August-Universit{\"a}t G\"ottingen, Friedrich-Hund-Platz 1, 37077 G{\"o}ttingen, Germany}
\address[id=UAD]{Center for Astrophysics and Space Science, NYUAD Institute, New York University Abu Dhabi, PO Box 129188 Abu Dhabi, UAE}

\runningauthor{Yang et al.}
\runningtitle{FARM}

\begin{abstract}
Maps of the magnetic field at the Sun's surface are commonly used as boundary conditions in space-weather modeling. However, continuous observations are only available from the Earth-facing part of the Sun's surface.
One commonly used approach to mitigate the lack of far-side information is to apply a surface flux transport (SFT) model to model the evolution of the magnetic field as the Sun rotates.
Helioseismology can image active regions on the far side using acoustic oscillations, and  hence has the potential to improve the modeled surface magnetic field. 
In this study, we propose a novel approach for estimating magnetic fields of active regions on the Sun's far side based on seismic measurements, and then include them into a SFT model.
To calibrate the conversion from helioseismic signal to magnetic field,  we apply our SFT model to line-of-sight magnetograms from Helioseismic and Magnetic Imager (HMI) on board the Solar Dynamics Observatory (SDO) to obtain reference maps of global magnetic fields (including the far side).  The resulting  magnetic maps are compared with helioseismic phase maps on the Sun's far side
computed using helioseismic holography. 
The spatial structure of the magnetic field within an active region is reflected in the spatial structure of the helioseismic phase shifts. 
We assign polarities to the unipolar magnetic-field concentrations based upon Hale's law and require approximate flux balance between the two polarities. 
From 2010 to 2024, we modeled 859 active regions, with an average total unsigned flux of $7.84 \cdot 10^{21}$~Mx and an average area of $4.48 \cdot 10^{10}$~km$^{2}$. Approximately $4.2\%$ of the active regions were found to have an anti-Hale configuration, which we manually corrected. Including these far-side active regions resulted in an average increase of $1.2\%$ (up to $25.3\%$) in the total unsigned magnetogram flux.
Comparisons between modeled open-field areas and EUV observations reveal a substantial improvement in agreement when far-side active regions are included.
This proof of concept study demonstrates the potential of the ``combined surface flux transport and helioseismic Far-side Active Region Model'' (FARM) to improve space-weather modeling.
\end{abstract}
\keywords{Active Regions, Models; Magnetic Fields, Photosphere; Helioseismology, Observations}
\end{opening}

\section{Introduction}
     \label{S-Introduction} 
     
Magnetic-field maps covering the whole solar surface are important for solar, heliospheric, and space-weather modeling. Our ability to obtain such maps is limited  as only about half of the solar surface can be directly observed from the Earth. 
So far, two approaches are frequently used in the space-weather community to construct full-surface magnetic-field maps: 1) the synoptic approach that stacks together bands of magnetic-field maps over one solar rotation to cover the entire surface of the Sun, and 2) the synchronic approach that evolves daily magnetograms with surface flux transport models. 
Both approaches have been demonstrated to work reasonably well, nevertheless, each of them has its own drawback. 

Synoptic maps suffer from “aging effects”: the magnetic field at different longitudes represent its evolution at different times  over a 27 day solar rotation \citep[e.g.,][]{2021Heinemann_farside}. The synchronic approach uses flux transport models to simulate changes of the magnetic field due to large-scale axisymmetric flows, such as meridional and rotational flows, and diffusion due to small-scale flows (e.g., ADAPT, \citealt{arge_10adapt} and AFT, \citealt{AFT2014}), but the
emergences of the active regions on the far side are either not modeled at all or not modeled on a regular basis.

The  direct solution to this problem is to send spacecraft to the Sun's far side to  obtain magnetograms. So far, the Photospheric and Magnetic Imager (PHI) \citep{solanki_20_o} on board the \textit{Solar Orbiter} (SO) \citep{mul_20so} is the only instrument that is capable of doing this. By using SO/PHI far-side magnetograms, \citet{PHI_impact_2024} demonstrated that a single missing active region on the far side can affect magnetic structure not only locally and but also globally. Due to the complex orbit of the SO mission, however, observations which cover a significant part of the Sun's far side are not available on regular bases. STEREO \citep[\textit{Solar TErrestrial RElations Observatories},][]{2008kaiser_STEREO} A and B are the other two spacecraft that have reached the Sun's far side, long before the SO mission.  Although no magnetogram is provided by STEREO, different authors have successfully demonstrated that some magnetic-field information can be derived from the SECCHI suite on-board \citep[\textit{Sun Earth Connection Coronal and Heliospheric Investigation},][]{2008howard_SECCHI}. \citet{2021Heinemann_farside} used $304~\AA$ filtergrams to empirically derive the open fields of far-side coronal holes, \citet{KIM2019} and \citet{JEONG2022} used  machine-learning technique to produce magnetic-field maps from EUV observations. Recent work of \citet{AFT_STEREO2024} included active regions on the far side to AFT by using 304~\AA \ filtergrams as proxies for magnetic fields, which showed a clear difference in the modeled open fluxes \citep{AFT_STEREO_test}. As a result of STEREO's orbit, there is only limited availability for far-side observations (roughly 2011-2015 at the time of writing). 

An alternative solution is to use solar acoustic oscillations observed on the Earth side to monitor magnetic activities on the Sun's far side. This method is known as helioseismic far-side imaging. It was proposed by \citet{LIN00b} to combine the solar oscillations in such a way to obtain maximum sensitivity to magnetized regions at focus points on the Sun's far side. Helioseismic  far-side imaging has been evidently proved to work by using chromospheric images from STEREO \citep[see, e.g.,][]{Liewer2014,ZHA19} and line-of-sight magnetograms from SO/PHI  \citep{yang_2023sophi}. This method has been used routinely to monitor the far side using observations of solar oscillations from Solar Dynamics Observatory (SDO)/Helioseismic and Magnetic Imager (HMI)\footnote{\href{https://jsoc.stanford.edu/data/farside/}{jsoc.stanford.edu/data/farside}} and 
National Solar Observatory (NSO)/Global Oscillation Network Group (GONG)\footnote{ \href{https://farside.nso.edu/}{farside.nso.edu}}.

Preliminary work by \citet{ARG13} demonstrated that helioseismic far-side imaging can improve the modeling of solar-wind speed. In their work, a bipolar  active region was identified manually  on seismic far-side maps and inserted into the ADAPT model, which improved modeled open fields. This is a very encouraging result, however, a decent amount of work is required to be able to include and update far-side active regions systematically and automatically. In this study, we develop the \textit{combined surface flux transport and helioseismic Far-side Active Region Model} in short FARM, a new full-surface magnetic-field map based on an example surface flux transport (SFT) model \citep[][]{Baumann2005}, which  automatically includes newly emerged active regions and updates existing ones on the solar far side. We use improved seismic far-side maps proposed by \citet{yang_2023method}, and convert active-region signals into magnetic fields using an empirical relation found in this work \citep[see also][]{GON07,yang_2023sophi}.

\section{Methods}
\subsection{Surface Flux Transport Model}
\label{sec_method_SFTM}

\begin{figure*}[!htb]
\begin{center}
\includegraphics[width=\linewidth]{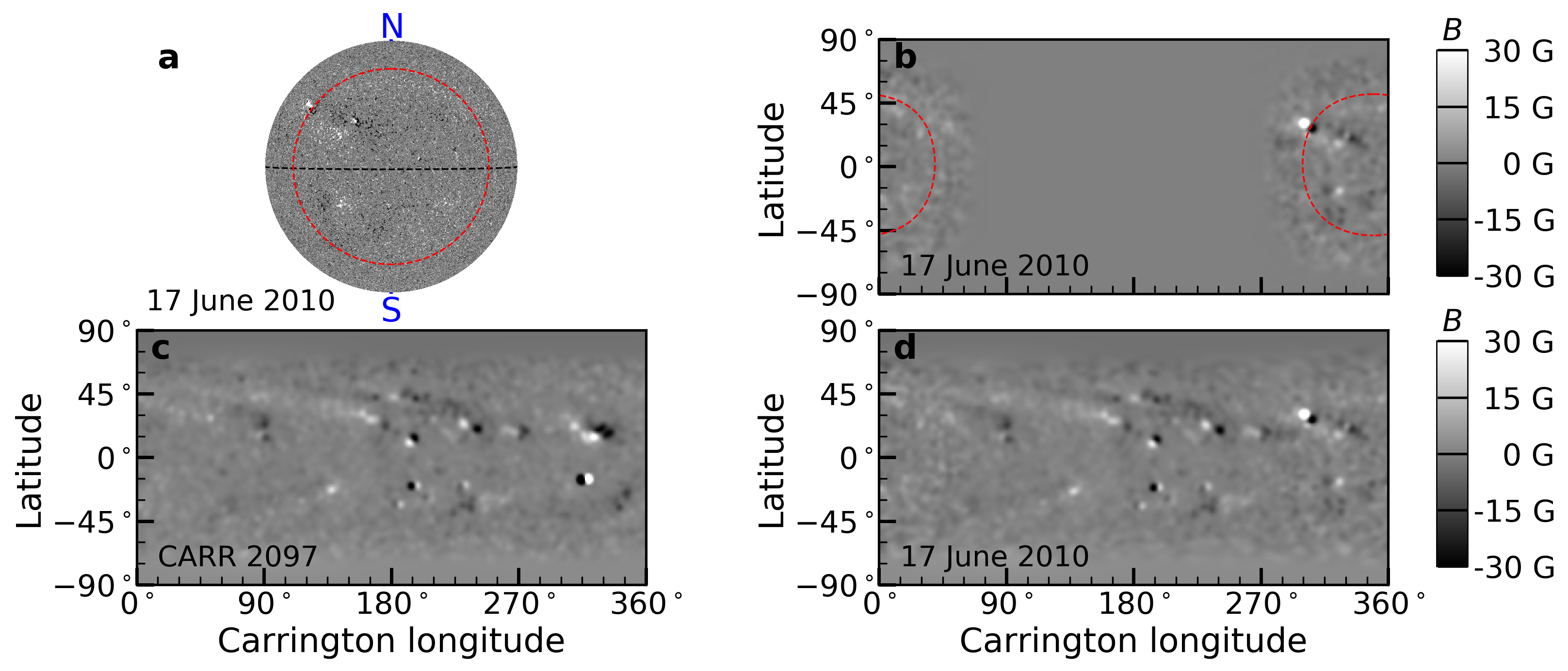} 
\caption{ Example magnetograms from SDO/HMI as inputs for the surface flux transport (SFT) model. \textit{Panel a}: 720~s line-of-sight magnetogram from SDO/HMI on  17 June 2010 as viewed by the camera. \textit{Panel b}: remapped magnetic fields in the Carrington reference frame using the  magnetogram as shown in panel a. \textit{Panel c}: full surface $B_r$ from JSOC/Stanford using synoptic maps from CARR 2097 (19 May -- 16 June 2010), which is used as the initial condition for our model. \textit{Panel d}: the resulting output magnetogram using SFT model (see texts in Section~\ref{sec_method_SFTM}). The red dashed curves on  panels a and b indicate the area 50$^\circ$  within the disk center.   
}
\label{fig.inputs}
\end{center}
\end{figure*}

Following  \cite{Baumann2005}, we assume  the photospheric magnetic field is radially oriented, which is denoted as $B_{r}$,   and is the solution to a passive scalar transport equation:
\begin{equation}
\begin{split}
    \frac{\partial B_r}{\partial t}& =   -\omega(\theta)\frac{\partial B_r}{\partial \varphi}-\frac{1}{\mathrm{R}_{\odot}\sin{\theta}}\frac{\partial} {\partial \theta}\big( v(\theta)B_r~\sin{\theta} \big) \\ & + \frac{\eta_h}{\mathrm{R}^{2}_{\odot}}\left[\frac{1}{\sin{\theta}}\frac{\partial}{\partial \theta}\big(\sin{\theta}\frac{\partial B_r}{\partial \theta}\big)+\frac{1}{\sin^{2}{\theta}}\frac{\partial^2 B_r}{\partial \varphi^{2}}\right]+ s(\theta,\varphi,t),
\end{split}  \label{eq.induction}
\end{equation}
where $\theta$ and  $\varphi$ denote  co-latitude and longitude in the Carrington reference frame. Here, $\mathrm{R}_{\odot}$ marks the solar radius,  $\omega$ and  $v$ are  differential rotation and  meridional flow, $\eta_h$ is an effective diffusion coefficient associated with \ non-stationary supergranular motions, and  $s$ is the source term.

For  $\omega$, we use a standard differential rotation profile from \citet{snodgrass_83rotation}:
\begin{equation}
\omega (\theta) = \frac{\left( 13.38 - 13.2 -2.3\cos^2\theta -1.62\cos^4\theta  \right) }{86400} \frac{\pi}{180} \ \mathrm{rad} \ \mathrm{s}^{-1}.
\end{equation}
For $v$, we take the state-of-the-art meridional flow model derived from helioseismology for Solar Cycle 23 \& 24, which takes inflows around active regions into account \citep[see][ and Figure~\ref{fig:mc} in the Appendix]{liang_18MC}:
\begin{align}
 v(\theta) & = v_{\mathrm{MC}}(\theta) + v_{\mathrm{inflow}} (\theta), \label{eq.MCst} \\
 v_{\mathrm{MC}}(\theta)  & =  \mathrm{V_{MC}} (-\frac{3 \sqrt{3}}{4}) \sin{2\theta}\sin{\theta},\\
 v_{\mathrm{inflow}}(\theta) & =  \mathrm{V_{inflow}}\bigg[ W_1 (\theta)\sin\left(\frac{9}{2} (\theta - \frac{33\pi}{180})\right) \sin^2\left(\frac{9}{4} (\theta - \frac{33\pi}{180})\right) \nonumber \\  
 &+ W_2 (\theta)\sin\left(\frac{9}{2} (\theta - \frac{67 \pi}{180})\right) \sin^2\left(\frac{9}{4} (\theta - \frac{67 \pi}{180})\right)  \bigg], \label{eq.MCed}
 \end{align}
where $v_{\mathrm{MC}}$ marks the unaffected meridional flow,  $v_{\mathrm{inflow}}$ is the inflow associated with active regions, and $\mathrm{V_{MC}}$ and $\mathrm{V_{inflow}}$ are the amplitudes set to be $14.2$  m~s$^{-1}$ and $10.9$ m s$^{-1}$. $W_1$ and  $W_2$ are two window functions 
\begin{equation}
W_1(\theta) =  \begin{cases}
      1  &   \qquad \mathrm{when} \quad \frac{33 \pi}{180} \leq \theta  \leq \frac{113 \pi}{180} ,   \\
      0 & \qquad \mathrm{otherwise}
   \end{cases}
\end{equation}
and 
\begin{equation}
W_2(\theta) =  \begin{cases}
      1  &   \qquad \mathrm{when} \quad \frac{67 \pi}{180} \leq \theta  \leq \frac{147 \pi}{180},   \\
      0 & \qquad \mathrm{otherwise}
   \end{cases}
\end{equation}
which define the range of the inflow.   

The SFT model models the evolution of field after emergence. For the near side, the field is directly observable.  We therefore use the 720~s line-of-sight magnetograms from \textit{SDO/HMI} \citep[][]{pesnell_2012SDO, schou_2012HMI} from 17 June 2010 to 1  July 2024 at cadence of one frame per day to
update the magnetic field on the Earth-facing side.
Each magnetogram is divided by the cosine of the angle between the local normal to the Sun’s surface and the line of sight to obtain an approximated $B_r$, and then remapped to the Carrington reference frame. A Gaussian filter is applied  to remove small scales which we do not aim to model. As the field is more reliable near the center of the disk, we apodize the data from 50$^\circ$ (red dashed curve in Figure~\ref{fig.inputs}a) till the limb. These processed magnetograms $B_r^{\mathrm{Obs.}}$ are updated daily in the SFT model (see Figure~\ref{fig.inputs}b for an example), with modeled magnetic fields $B_r$ being replaced by $W_a B_r^{\mathrm{Obs.}} + (1 -  W_a) B_r$, where $W_a$ is the smoothing window function used for apodization.

The diffusion coefficient $\eta_h$ is set to be $250$ km$^2$ s$^{-1}$ to match the amplitude of polar field from observations \citep[see Figure~\ref{fig:polar_field} in the Appendix and][]{CAM10}.
We solve Equation~\ref{eq.induction} in the spectral space with harmonic degrees up to 80 using the code developed by \cite{Baumann2005}. 
We use the synoptic radial magnetic fields (JSOC series: \textsl{hmi.synoptic\_mr\_polfil\_720s}\textrm{}) of Carrington rotation 2097  as the initial $B_r$ (see Figure~\ref{fig.inputs}c). 
 
Thus far the code produces full surface magnetograms without using seismic far-side images. The magnetic field on the far side is assumed to evolve according to Equation~\ref{eq.induction}. 
To avoid changing far-side magnetic fields when assimilating the front-side data, we do not enforce a zero total flux in the SFT model. The total flux fluctuates between $\pm10\%$ of the total unsigned flux in the simulation. 

\begin{figure}[!htb]
\begin{center}
\includegraphics[width=\linewidth]{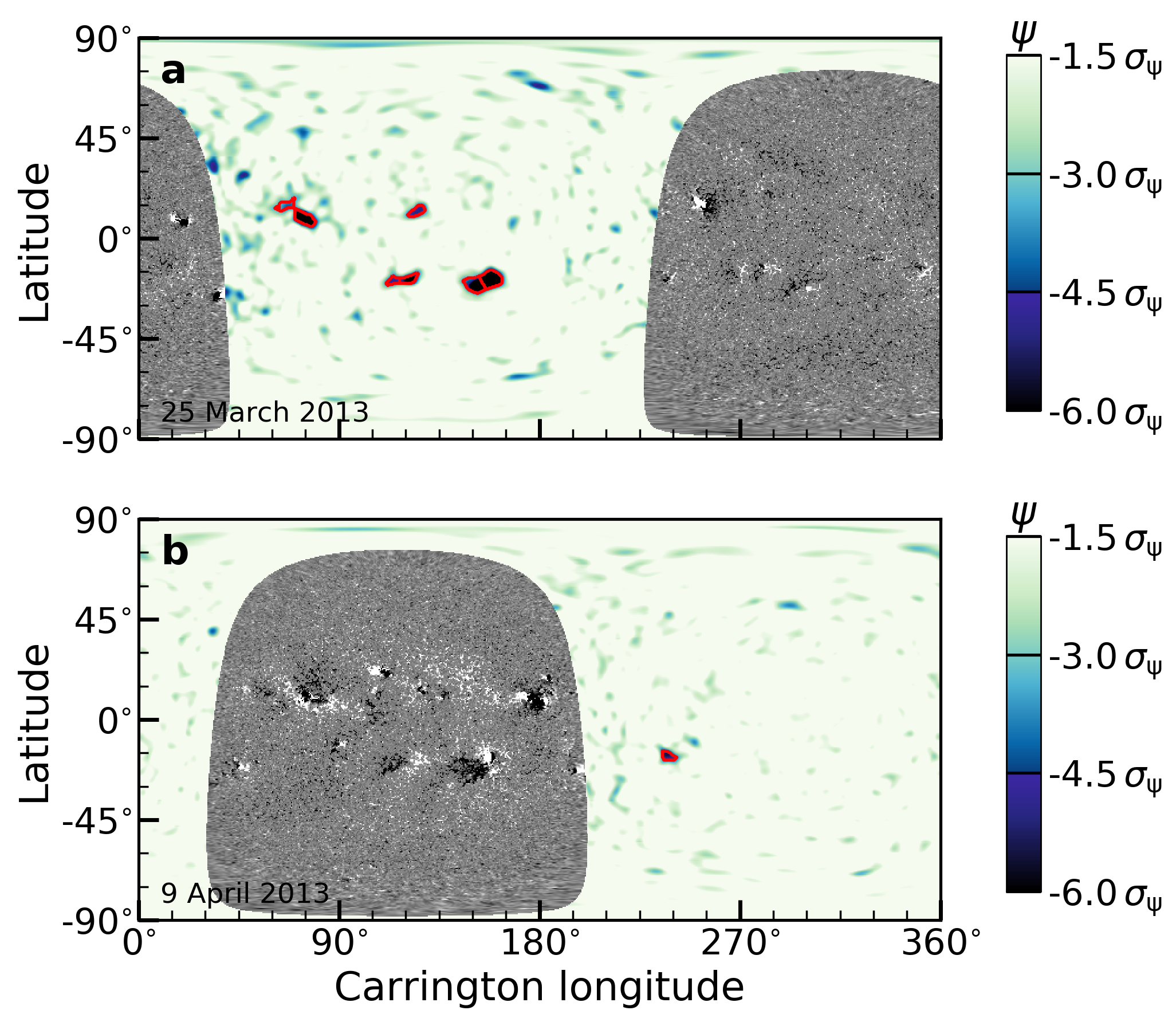} 
\caption{Example helioseismic far-side images in Carrington reference frame. Line-of-sight magnetograms from SDO/HMI shown are on the front side in black and white in the range of -30 to 30 G. Seismic images are shown in the range of $-6 \sigma_\psi$ to $-1.5 \sigma_\psi$, where $\sigma_\psi$ is the noise as measured from the quiet-Sun area in each map.  Panels a and b are example maps on 25 March 2013 and  15 days later on 9 April 2013. In both panels, active regions detected by helioseismology are indicated by red contours.  
}
\label{fig_farside_full}
\end{center}
\end{figure}

\subsection{Seismic Far-side Images}

Seismic far-side images are obtained by applying the method as described by \citet[][]{yang_2023method} to Dopplergrams from SDO/HMI. Specifically, overlapping segments of 31~hr Dopplergrams are used to compute maps of seismic phase shifts $\psi$ at a cadence of 12~hr. We average five consecutive segments to obtain 79~hr seismic maps, and use these maps to detect active regions on the far side. All seismic maps are in the Carrington reference frame, with a spacing of $0.5^\circ$ in both latitude and longitude directions. We smooth the maps by a Gaussian kernel to remove spatial scales smaller than the resolution limit. For an example see Figure~\ref{fig_farside_full}.

\begin{figure*}[!htb]
\begin{center}
\includegraphics[width=\linewidth]{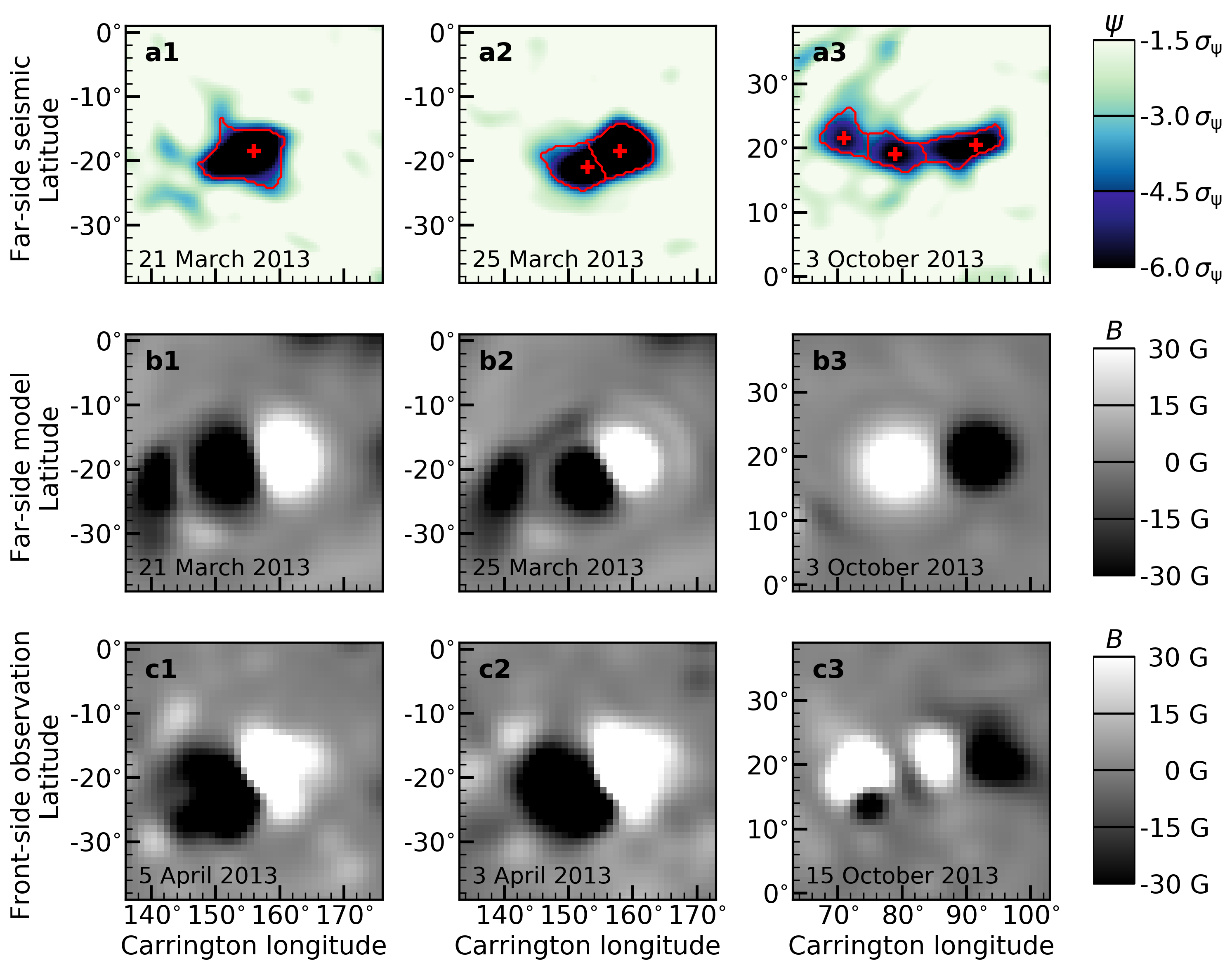} 
\caption{Example modeled active-region magnetic fields using helioseismic far-side maps. The top panels a$1$ to a$3$ show zoom of seismic phase maps around active regions with 1 to 3 poles. The middle panels b$1$ to b$3$ show corresponding modeled magnetic fields using seismic phase maps as seen on the top panels. The bottom panels c$1$ to c$3$ show the magnetograms when the active regions appear on the Earth's side of view.    
}
\label{fig_model_br}
\end{center}
\end{figure*}

\begin{figure*}[!htb]
\begin{center}
\includegraphics[width=\linewidth]{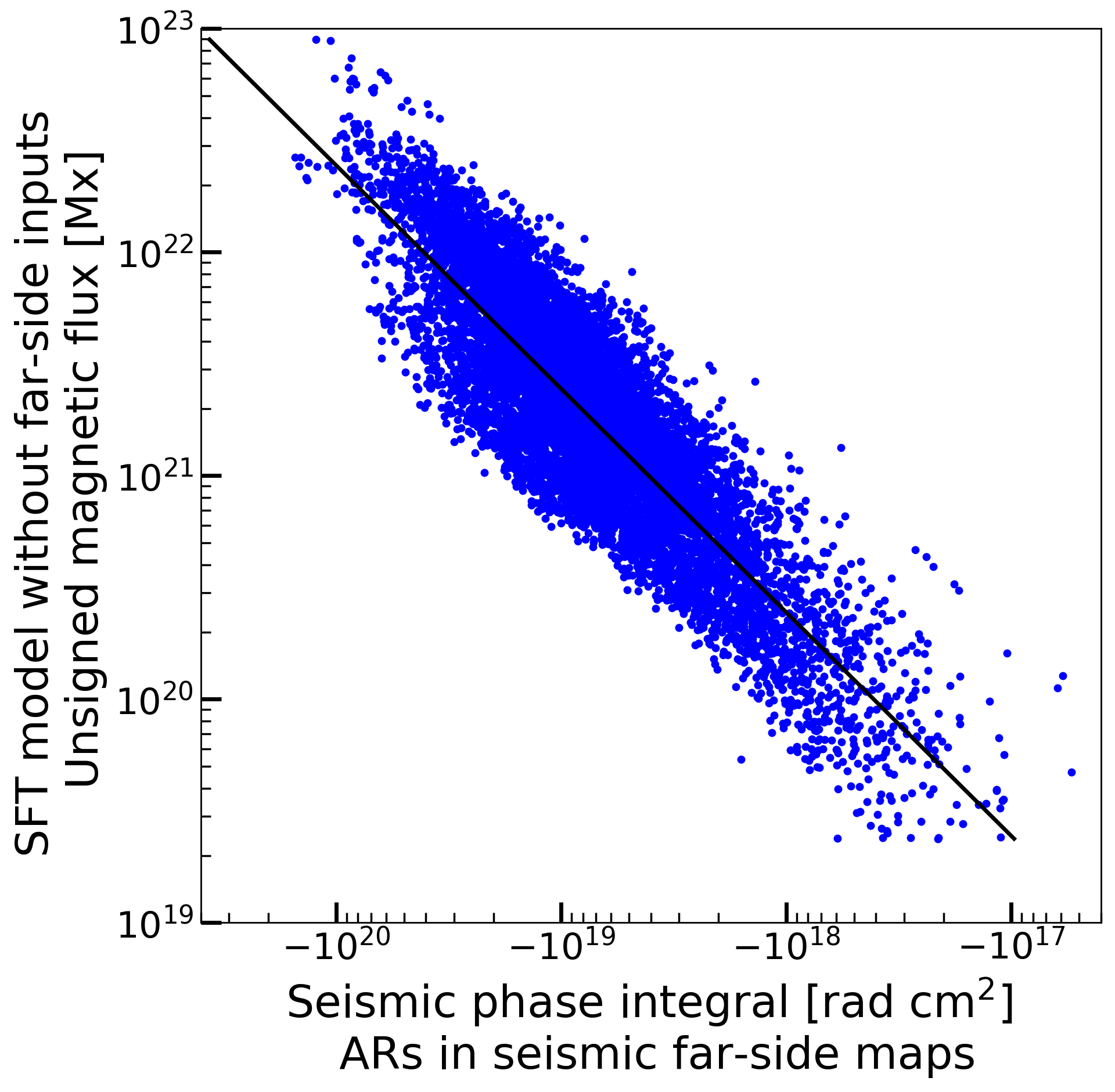} 
\caption{Unsigned magnetic flux as a function of integrated seismic phase shifts  from the individual pole in each active region. Magnetograms from the SFT model (without far-side inputs) are used to calculate the unsigned magnetic flux.  We assume that the  unsigned magnetic flux of the active region is proportional to the  integrated seismic phase across the active region (see Equation~\ref{eq.flux_relation}), and the black curve shows the best fit with this assumed form.} 
\label{fig_f_relation}
\end{center}
\end{figure*}

\subsection{Locating Active Regions on the Far Side using Seismic Maps}
\label{sec.label_AR_seismic}
We aim to include active region emergence in the SFT model (smaller emergences are not reliably detectable using far-side imaging with seismology).   
To locate far-side active regions, we  create a binary mask based on a threshold of three times the seismic noise and bin each binary mask by a factor of $5\times 5$ in longitude and latitude (to $2.5^\circ$). 
This method selects patches resemble the active-region area as visually identified on STEREO 304~\AA\ images. 
Because regions with sizes less than 800~$\mu\textrm{hs}$ are often false detections due to noise,  we discard them.

A single active-region patch identified using the above method will usually contain more than one pole (magnetic feature). To separate the individual poles, we identify the local minima within each patch as the center of a pole (see red crosses in Figures~\ref{fig_model_br} (a1-a3)). To determine which pixels in a feature belong to which pole, 
we first find the shortest path between each pair of poles which lies entirely in the active region patch. We then find the local maxima along this path (where the seismic phase will be closest to zero). We finally separate the two poles by the line through the local maxima which is normal to the path.  For an example see Figures~\ref{fig_farside_full} and ~\ref{fig_model_br}a.  This procedure was implemented using \textsl{scipy} and \textsl{skimage}.

\subsection{An Empirical Relation Between Seismic Signals and Magnetic Fields}

\citet{yang_2023sophi} report that a simple linear relation exists between mean seismic phase shifts and mean unsigned magnetic field within active regions. 
In this work, we are interested in how to go from the seismic phase shifts integrated over identified active regions to the amount of unsigned magnetic flux in each active region.  
To obtain a relation between unsigned magnetic flux  and seismic measurements, we restrict ourselves to active regions which were observed on the disk before their detection on the Sun's far side. We enforced this constraint by requiring a mean field strength  of at least 17~G over active regions in the SFT model. In particular, active-region areas are determined by seismic maps, and this constraint is applied at the time of comparison on the Sun's far side.
We assume that on average the SFT model reproduces their evolution well enough for the SFT model to reproduce the amount of unsigned magnetic flux.

Figure~\ref{fig_f_relation} shows the unsigned magnetic flux  $\int_{P} |B_r| dS$  as a function of integrated seismic phase shifts  $\int_{P} \psi dS$ from the individual pole $P$ in each  active region. The individual pole $P$ is determined by seismic maps using the method described in Section~\ref{sec.label_AR_seismic}. 
We assume a linear relation between these two quantities
\begin{equation} 
\int_{P} |B_r| dS = C \int_{P} \psi dS
\label{eq.flux_relation}
\end{equation}  
and find
$C = -245 $ G rad$^{-1}$.

\subsection{Modeling Active-region Magnetic Fields}
We now have the number and positions of the poles in an active region, the integrated seismic phase shift of each pole, and a relation to convert the integrated phase shifts into magnetic fluxes. 
We now describe how we model magnetic fields from seismic phase shifts.\

\begin{figure*}[!htb]
\begin{center}
\includegraphics[width=0.9\linewidth]{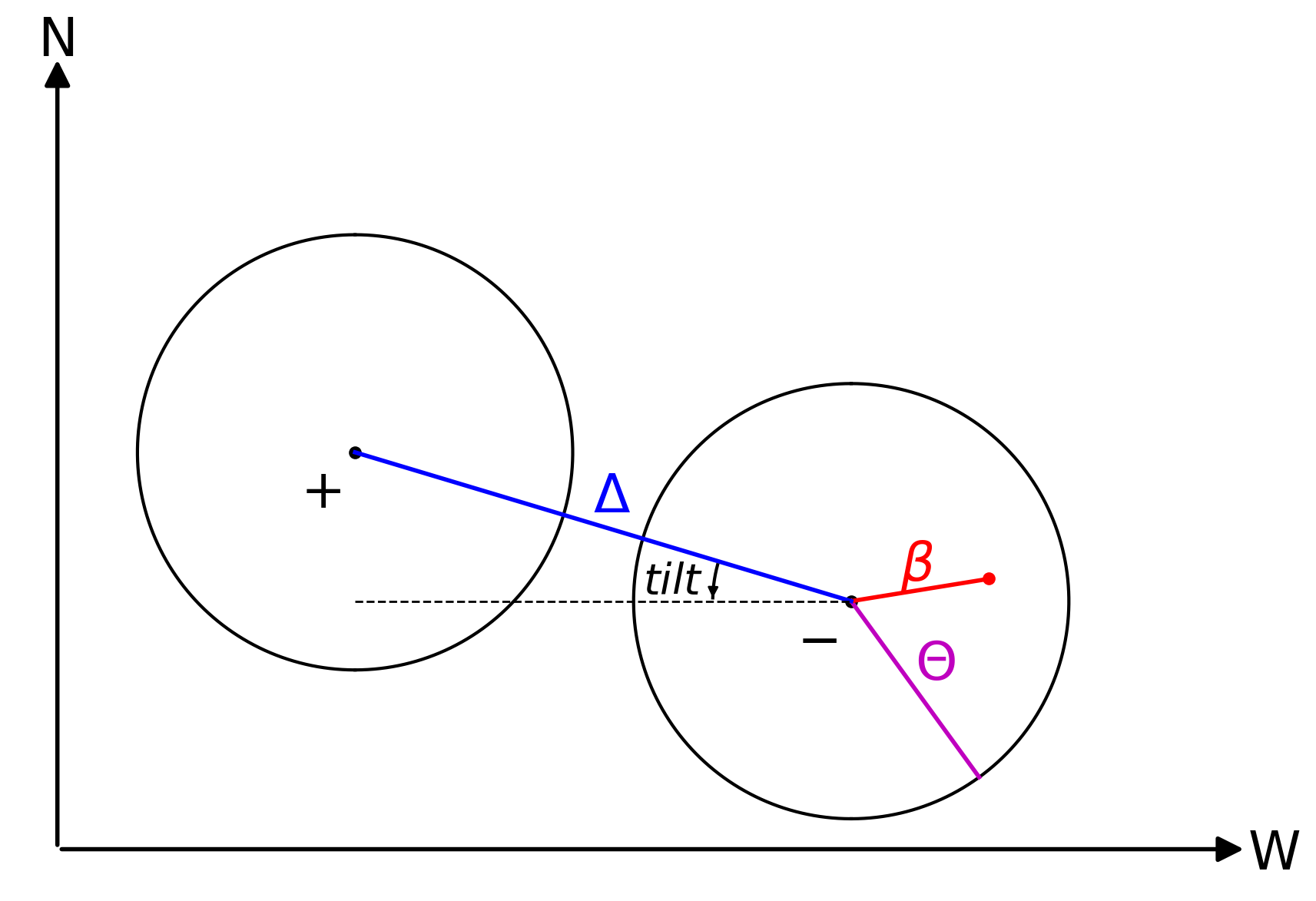} 
\caption{A simple geometry plot of a bipolar active region which has a positive tilt  angle in the
northern hemisphere. The two circles indicate the leading (most westward) and the trailing polarities. The leading polarity is negative during even-numbered solar cycles (e.g., Solar Cycle 24) and is closer to the equator. $\Delta$ is the angular distance between the centers of the two polarities (black dots), $\Theta$ indicates the angular radius of a given polarity, and $\beta$ marks the angular distance between the center of the polarity and an arbitrary point (red dot).}
\label{fig.geo_demo}
\end{center}
\end{figure*}

If we have one isolated pole, i.e., no distinct poles as determined by helioseismology, we modeled it as a bipolar active region by assuming that one polarity is too weak to be detected. The modeled region is centered at the center of the detected area, and the separation between two polarities $\Delta$ follows \citet{CAM10},
\begin{equation}
\Delta = \frac{0.45 \ S^{\frac{1}{2}} }{ \mathrm{R_\odot}} \ \mathrm{rad},\label{eq.sepration_biploar}
\end{equation}
where $S$ is the total area of the bipolar region.
The tilt angle follows \citet{wang_91tilt}
\begin{equation}
tilt = \mathrm{sgn}(\lambda_0) \ \mathrm{arcsin}\left(0.48 \ \mathrm{sin}(|\lambda_0|)  + 0.03\right)   \label{eq_tilt_mono},
\end{equation} 
where $\lambda_0$ is the latitude for the center of the region. The sign of the latitude $\lambda_0$ 
 is added such that active regions which obey the Joy's law will  have positive/negative tilt angles in the northern/southern hemisphere. For a sketch of the geometry see Figure~\ref{fig.geo_demo}.

If the active region contains two or three poles then we model them as bipolar active regions.  The polarity for modeled bipolar regions are assigned according to Hale's law, that is, the polarity of the leading (most westward) pole in the active region is given the polarity of leading spots for that hemisphere and cycle. The trailing (most easterly) pole is given the opposite (trailing) polarity. In this work, we consider 15 December 2019 as the start of Solar Cycle 25.  For active regions which contain three poles, we first consider the two poles with the largest integrated seismic phase shift. We determine the polarity of these two poles using Hale's law as we would for a bipolar active region. We then attribute the integrated seismic phase shift and area of the third pole $p_3$ to the pole with the shortest distance to it, and discard $p_3$ in later analysis.

We now impose the constraint that the total flux of the active region should be zero. We regard the magnetic fluxes of each pole based on the integrated phase shift as an estimate of the true flux of each pole. The constrain is exact. 
The problem of finding the true fluxes is then overdetermined. We find the fluxes which exactly obey the constraint and minimizes the difference in the L2 sense from the values inferred from the helioseismic phase shifts.

We assume the radial magnetic field of each pole $B_p$ is distributed in space according to
\begin{equation}
B_p =  A \exp\left\{\frac{ 2 (\mathrm{cos \beta} - 1)}{\delta^2}\right\}, \label{eq.br_model}  
\end{equation}
where $\beta$ is the angular distance to the center of the pole,  $\delta$ determines the width, and $A$ is determined by the total flux of the pole. We define an effective circular region $P_\mathrm{effective}$, which has the same area and center location as the pole. $\delta$ is taken to the angular radius $\Theta$ of  $P_\mathrm{effective}$. Limited by the maximum harmonic degree 80 used in the SFT model, 
if $\delta <$ 4$^\circ$ we replace it by  $\delta=$ 4$^\circ$ for $\delta <$ 4$^\circ$\citep[see also][]{CAM10}.

Figure~\ref{fig_model_br} shows examples of modeled $B_p$ for active regions with 1 to 3 poles (as indicated by the numbers on the top left in each panel). The top panel (Figures~\ref{fig_model_br}~(a1-a3)) shows the seismic maps with detected active regions (red contour), and the middle panel (Figures~\ref{fig_model_br} (b1-b3)) shows the modeled $B_p$. We follow these active regions till they come to the Earth's view and show the corresponding $B_r$ from SDO/HMI in the bottom panel (Figures~\ref{fig_model_br} (c1-c3)).

\subsection{Far-side Sources as a Function  of Time}

For each pole, a corresponding source term is added
\begin{eqnarray}
 s_p(\theta,\varphi,t) &=& \big(B_p(\theta,\varphi,t_e) - B_r(\theta,\varphi,t_e-\Delta t)\big) \delta_{t, t_e},
\end{eqnarray}
where $\Delta t$ is the time step used in the SFT model (1 day). $B_p$ is the modeled active-region magnetic field based on seismic maps, which is assumed to be zero at points that are 2~$\delta$ away from the center of the pole; $B_r$ represents magnetic field from the SFT model.  $\delta_{t, t_e}$ is the Kronecker delta, which is one when $t = t_e$ and zero otherwise.
$t_e$ is such that the unsigned magnetic flux of $B_p(\theta,\varphi, t_e)$ is $15\%$ higher than that of $B_p(\theta,\varphi, t_e-\Delta t)$ and the coinciding SFT model without far-side inputs at $t_e$. All these fluxes are calculated and compared before running the SFT model with far-side sources.
When an active region already exists in the SFT model, it often has a larger area than that of $B_p$, as the magnetic field can only decay in the model. This extended area will lead to excess fluxes in the SFT model. To avoid this problem, we first find the area  $P_\mathrm{large}$, where $|B_r|>5$~G  and has overlap with $B_p(\theta,\varphi,t_e)$ in the SFT model. We then replace $B_r$ within  $P_\mathrm{large}$ by the standard deviation of quiet-Sun values, which is obtained by using $B_r$ in the SFT model at each time step for $|B_r|\leqslant5$~G. A minus sign is given at points with negative polarity in the SFT model before including the far-side source.
This cleaning procedure is applied only when the unsigned flux of  $B_p$ from $P_\mathrm{large}$ is 4 times larger than the unsigned flux of $B_r$ from the rest area in $P_\mathrm{large}$.

\subsection{Anti-Hale Active Regions}
As previously stated, a major drawback of using helioseismic maps to derive far-side active regions is the lack of polarity information. Additionally, although we can estimate the polarity and pole-configuration using Hale's law and Joy’s law, this does not account for all active regions. Active regions may appear whose leading polarity is the opposite of the one predicted by Hale's law. These are known as ``parasitic'' or ``anti-Hale'' active regions \citep[e.g.,][]{wang_89spot}. Our model is not capable of distinguishing anti-Hale active regions, especially without magnetic-field information when the active region rotates into Earth's field of view. We manually compared the polarities of the far-side active regions to the polarity of the active regions at same position (in the \textit{Carrington} frame after correcting for differential rotation) when in the disk center as seen from Earth. This allowed us to identify 36 modeled active regions with incorrect polarity, accounting for $4.2\%$ of all modeled ARs. 

Despite the efforts invested in correcting the polarity for the anti-Hale regions, not all modeled active regions could be matched with a respective active region in the magnetic-field data on the Earth-facing side. Reasons for this include active regions that have dissolved before the Sun rotated to the front side and configurations that are too complex for a clear association to be made. A total of 26 active regions, or $3.0\%$ of the modeled active regions, fell into this category. As we were unable to clearly identify them as artifacts or having the wrong polarity, they remain in the model.

\begin{figure}
\centering
\includegraphics[width=\linewidth]{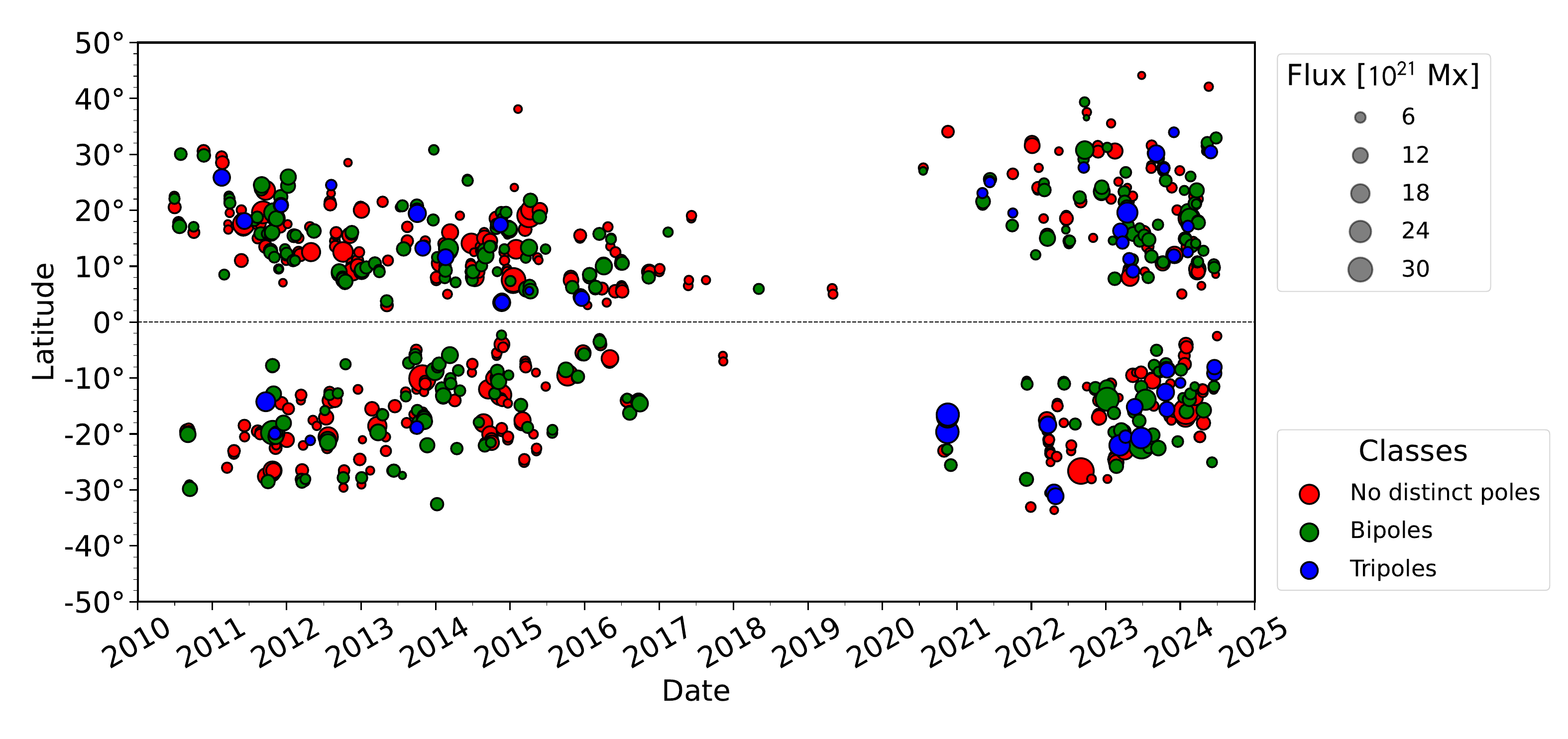}
  \caption{Distribution of far-side active regions included in the model as a function  of time, latitude, unsigned magnetic flux, and configuration. The y-axis shows the position of the geometric center of mass (latitudinal component only) and the x-axis shows the time when each active region that was detected with far-side helioseismology and consequently included into the SFT model. The size displays the total unsigned flux and the color indicates the pole configuration.} 
     \label{fig:fluxdistr}
\end{figure}

\begin{table*}
\caption{Far-side active regions included into the SFT model. Area and Total flux (unsigned) are average values of the respective categories.}             
\label{tab:stats}      
\centering       
\renewcommand{\arraystretch}{1.5}
\begin{tabular}{l | l  c c c c }     
\hline      
Seismic observations & Model & Number & Percentage & Area & Total Flux \\ 
 &  & [N] & [$\%$] & [$10^{10}~$km$²$] & [$10^{21}~$Mx] \\ \hline
No distinct poles & Bipole  &  $493$  &  $57.39$  &  $4.18$  &  $7.30$\\
Bipole & Bipole  & $312$  &  $36.32$  &  $4.52$  &  $8.13$ \\
Tripole & Bipole  &  $54$  &  $6.29$  &  $7.11$  &  11.11 \\ \hline
&Total & $859$  &  $100.00$  &  $3851.69^{*}$   &  6735.08\\
\hline

\end{tabular}
    {\footnotesize \parbox{0.8\linewidth}{$^{*}$ This value is more than six times the Sun's total surface area.}}
\end{table*}

\section{Statistics}
Over a period spanning $\sim$14 years, from 17 June 2010 to 1 July 2024, we included 859 far-side located active regions, which are active regions that either emerged on the far side or displayed an increase in flux of at least $15\%$ compared to existing active regions at same locations. A single active region might be included multiple times at consecutive time steps if the growth conditions are met. The area of the inserted active regions varies between $3.16$ to $12.65 \cdot 10^{10} \mathrm{km}^{2}$, with a mean area of $4.48 \cdot 10^{10} \mathrm{km}^{2}$. The mean total unsigned flux of the modeled regions is $7.84 \cdot 10^{21}\mathrm{Mx}$, ranging from $0.99$ to $35.95 \cdot 10^{21} \mathrm{Mx}$. Table~\ref{tab:stats} lists the fraction of far-side sources  for each respective magnetic configuration, along with the associated average total unsigned flux and area. The average time between a far-side region to appear in our SFT model and the time it rotates to the disk center of Earth's field of view is $13.99 $ days, which aligns with our expectations.

The distribution of the inserted active regions as a function of time, latitude, total unsigned magnetic flux, and magnetic configuration (number of poles) is depicted in Figure~\ref{fig:fluxdistr}. Far-side sources follow sunspot  butterfly diagrams, with active regions appearing at higher latitudes during the rising phase and primarily closer to the equator during other phases. Additionally, during solar minimum (2017-2021), only a few far-side active regions were detected and included. The most complex active region configurations tended to appear preferentially during solar maximum, particularly towards the latest ongoing maximum around late 2022 to 2024. This observation indicates that active regions emerging or growing on the far side contribute significantly to the total flux. Throughout the SDO-era, FARM magnetograms, on average, contain a $1.3\%$ higher total unsigned flux than predicted by the model without the far-side emergences, but can contain up to $23.2\%$ more flux at the time when far-side active regions are included. During periods of high solar activity, the average contribution increases to $1.5\%$ and $1.6\%$ for 2011-2016 and 2022-2024, respectively (also refer to Figure~\ref{fig:excessflux} in the appendix). During the period of low solar activity between 2018 and 2021, the total unsigned flux in FARM and the SFT model without far-side sources is virtually equal due to the absence of emerging active regions.
\begin{figure}

\centering
\includegraphics[width=1.0\textwidth]{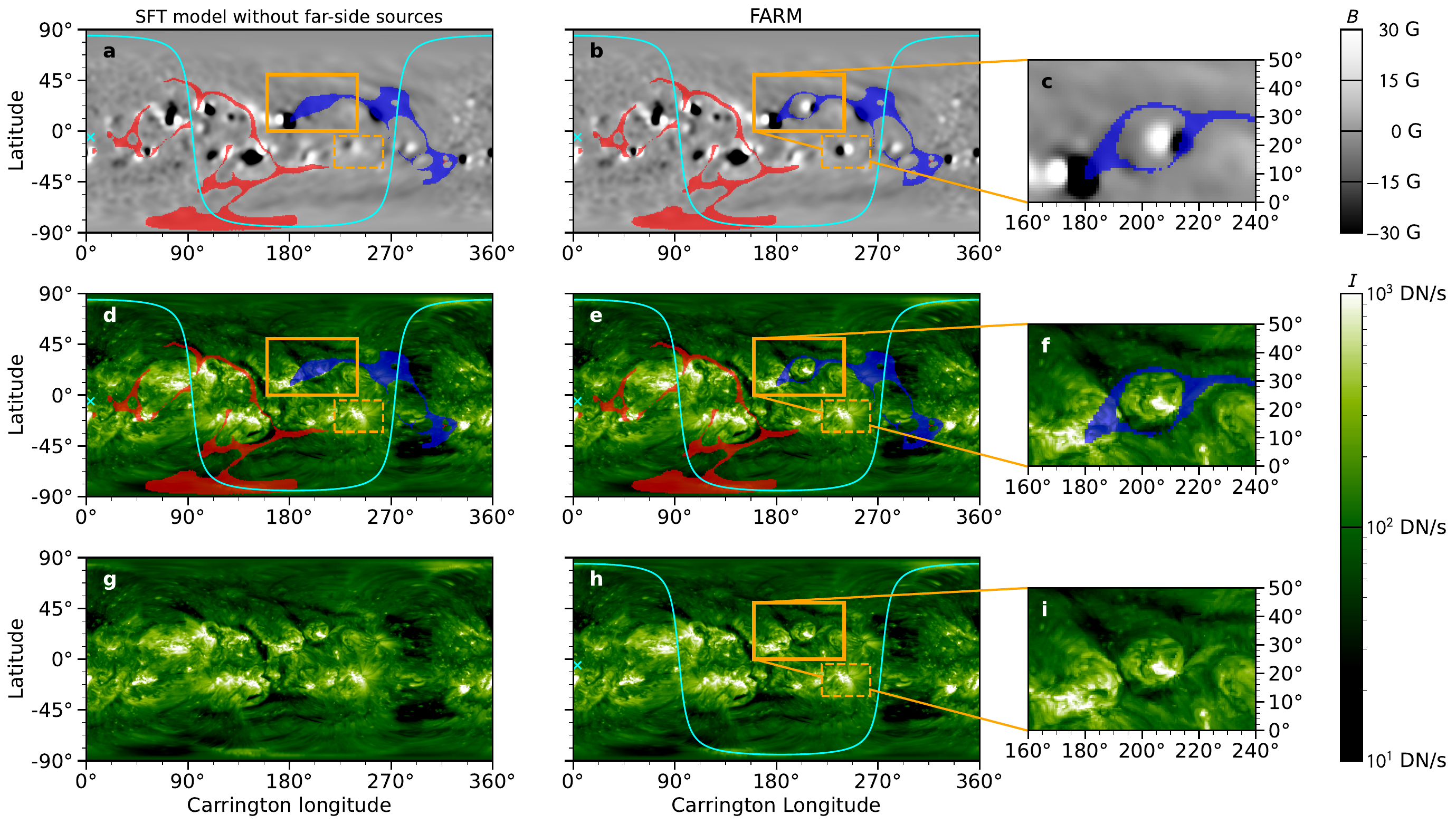}
  \caption{Magnetograms and combined EUV filtergrams with overlaid PFSS model results from the 17 April 2013 12:00UT. Panels \textbf{a} and \textbf{b} show  the SFT model without far-side inputs and FARM magnetograms with their respective open-field regions overlaid. Panels \textbf{d} and \textbf{e} show combined EUV synoptic charts (provided by Predictive Science Inc. \href{https://q.predsci.com/CHMAP-map-browser/}{q.predsci.com/CHMAP-map-browser}) that combine SDO/AIA 193~\AA\ and STEREO 195~\AA\ images for a synchronous $360^{\circ}$ map of the Sun. The open-field maps from the SFT model without far-side source and FARM are overlaid. The red and blue shades show the positive and negative polarity footpoints of the open field respectively. The orange rectangles mark areas of interest where active regions were included into the magnetogram. The cyan line shows the field of view from Earth, and the cyan ``x" indicates the disk center. Panels \textbf{g} and \textbf{h} show the combined EUV synoptic charts only without overlaying open-field maps. Panel \textbf{c}, \textbf{f}, and \textbf{i} show the zoom of the region as indicated by the orange rectangles in panels \textbf{b}, \textbf{e}, and  \textbf{h}.} 
     \label{fig:openclosed}
\end{figure}

\section{Conclusions and Outlook}
This work presents a way to convert seismic far-side signals into magnetic fields. 
Seismic phase maps on the far side are used to infer unsigned magnetic fluxes. Different magnetic parts (poles) of an active region are define by the spatial structure (local minima) in the helioseismic phase shifts.  Hale's law is then used to assign a polarity to each of the magnetic features in each active region.

The resulting magnetic fields are used to improve surface flux transport (SFT) model applied to SDO/HMI magnetograms. 
Figure~\ref{fig:openclosed} shows example open field modeled by using the SFT model with and without the inclusion of the far-side emergences. To avoid spurious electric fields, the flux is multiplicatively balanced before performing the open-field calculations. Specifically, the two polarities are rescaled linearly  in such a way that the total flux is zero while conserving the total unsigned flux. A clear improvement is seen in terms of the open fields with FARM magnetograms and STEREO 195~\AA\ images.
Solar Orbiter, now operational, measures the Sun's surface magnetic fields from various Earth-Sun-S/C angles including the far side. This offers an excellent opportunity to validate and calibrate FARM.  Particularly, these measurements can be compared to seismic maps to clean the image and to obtain more accurate relation between seismic phase shifts and magnetic fields to improve the accuracy of FARM magnetograms.

This proof of concept study demonstrates that FARM magnetograms have the capability to significantly improve solar wind modeling, hence produce more accurate heliospheric magnetic fields for space-weather forecasting.

\begin{acknowledgments}
We thank an anonymous referee for useful comments.
The SDO and STEREO image data are available by courtesy of NASA and the respective science teams. 
\end{acknowledgments}

\begin{authorcontribution}
D.Y. designed research and implemented the model. S.G.H. identified anti-Hale active regions, performed the statistical analysis, and carried out open-field calculation. All authors discussed the results and contributed to the final manuscript.
\end{authorcontribution}

\begin{fundinginformation}
 D.Y., R.H.C. and L.G. acknowledge support from ERC Synergy grant WHOLE SUN 810218. S.G.H. acknowledges funding by the Austrian Science Fund (FWF): Erwin-Schr\"odinger fellowship J-4560. L.G. acknowledges support from Deutsche Forschungsgemeinschaft (DFG) grant SFB 1456/432680300 Mathematics of Experiment (project C04) and from the NYUAD Center for Astrophysics and Space Science. 
\end{fundinginformation}

\begin{dataavailability}
Data generated during the current study are available from the corresponding author on reasonable request.
\end{dataavailability}

\bibliographystyle{spr-mp-sola}
\bibliography{bib}  

\appendix

\section{Addional Figures}
\begin{figure}[!htb]
\begin{center}
\includegraphics[width=\linewidth]{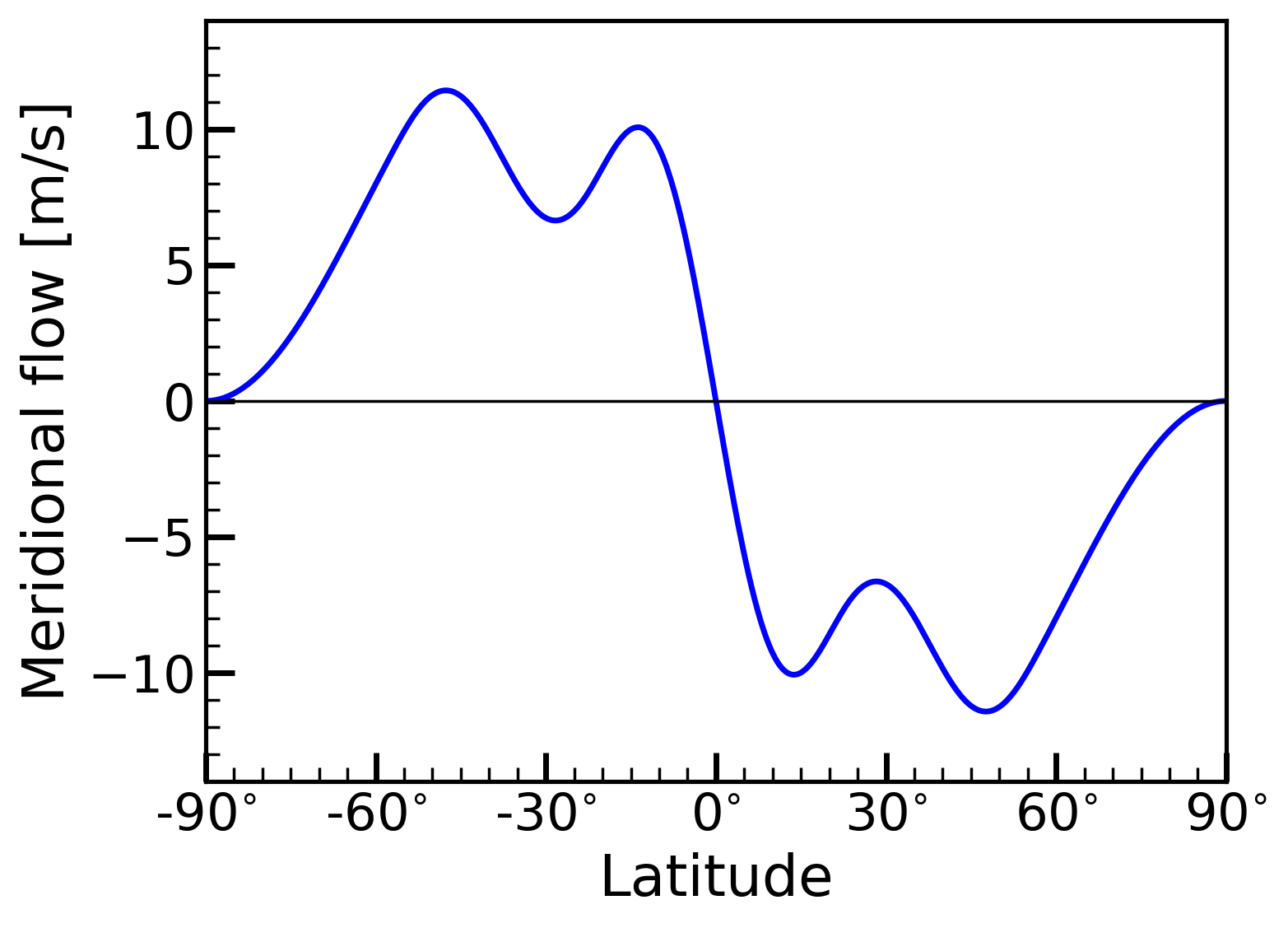} 
\caption{Meridional flow model proposed by \citet{liang_18MC} for Solar Cycle 23 \& 24 (see, Equations~\ref{eq.MCst} to \ref{eq.MCed}). Inflows around active regions were taken into account.}
\label{fig:mc}
\end{center}
\end{figure}

\begin{figure}[!htb]
\begin{center}
\includegraphics[width=\linewidth]{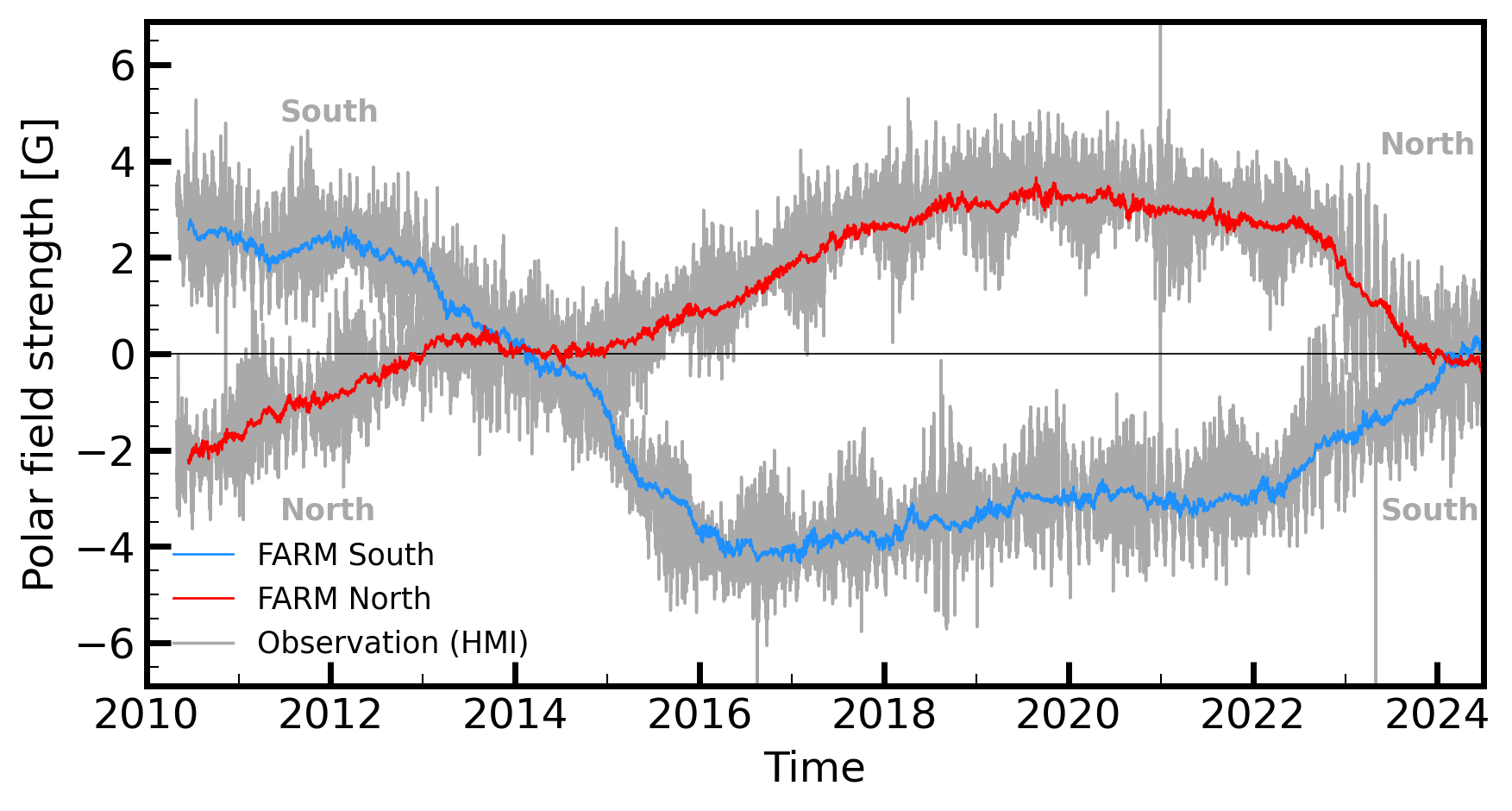} 
\caption{Comparisons of the polar field strength between FARM magnetograms (blue and red) and observations from SDO/HMI (gray). The polar field strength from FARM is calculated by taking the mean of the magnetic fields from $\pm$60$^\circ$ to $\pm$90$^\circ$. For HMI observations, we use the polar field strength provided by the JSOC (series name \textsl{hmi.meanpf\_720s} at a 12hr cadence). }
\label{fig:polar_field}
\end{center}
\end{figure}

\begin{figure}
\centering
\includegraphics[width=\linewidth]{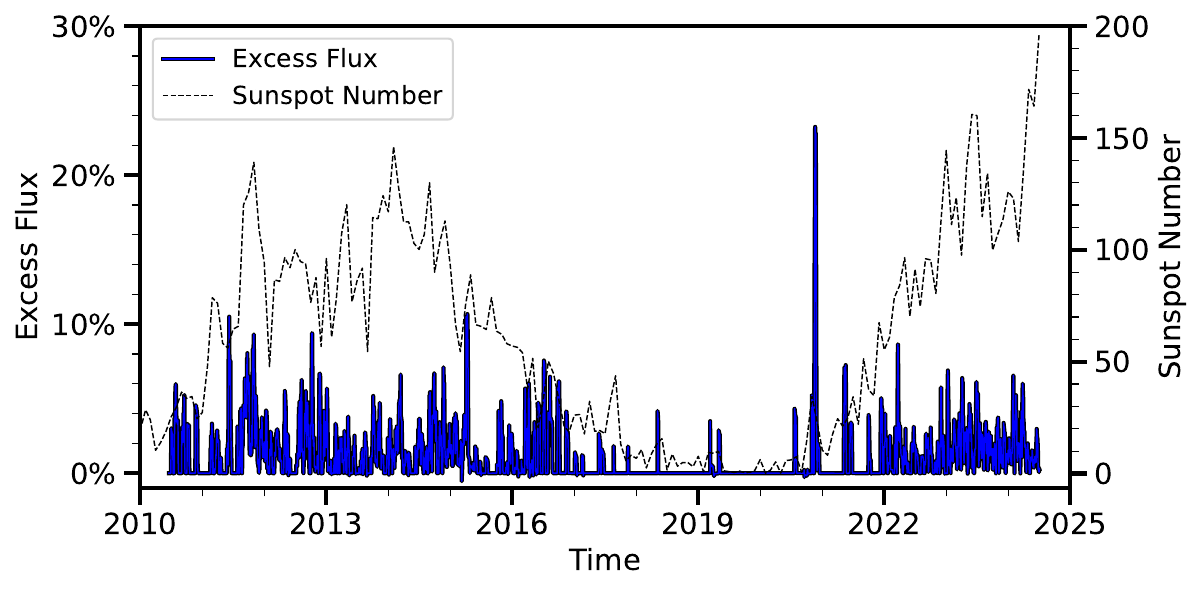}
  \caption{Percentual excess of total unsigned magnetic flux and monthly sunspot number from the SILSO World Data Center \citep{sidc} over a full solar cycle. We define the excess flux as the ratio of the total unsigned flux calculated from FARM divided by the respective flux of the SFT model without far-side inputs. }
     \label{fig:excessflux}
\end{figure}

\end{article} 

\end{document}